\documentclass[11pt]{article}

\usepackage[preprint]{acl}

\usepackage{times}
\usepackage{latexsym}

\usepackage[T1]{fontenc}

\usepackage[utf8]{inputenc}

\usepackage{microtype}

\usepackage{inconsolata}

\usepackage{graphicx}
\usepackage{epstopdf}
\usepackage{multirow}
\usepackage{enumitem}
\usepackage{amsmath}
\usepackage{booktabs}
\usepackage{adjustbox}
\usepackage{array}
\usepackage{caption}

\newcommand{\model}{PROOF}
\newcommand{\modelbf}{\textbf{PROOF}}

\title{From Reading Code to Reading Spec: A Verified Layer for LLM-Driven Codebase Maintenance}

\author{
  \textbf{Xinhao Zhang}\textsuperscript{1}, \
  \textbf{Jingjie Lu}\textsuperscript{1}, \
  \textbf{Kunpeng Liu}\textsuperscript{2}, \ and \
  \textbf{Fei Xie}\textsuperscript{1}\thanks{\ Corresponding author.}
\\
\\
  \textsuperscript{1}Portland State University, USA \\
  \textsuperscript{2}Clemson University, USA
\\
  \texttt{\{xinhaoz, jingjie, xie\}@pdx.edu}, \texttt{kunpenl@clemson.edu}
}

\begin{document}
\maketitle
\begin{abstract}
The rapid growth of LLM-generated code increases software complexity and the maintenance burden on engineers. While LLMs offer a potential automated alternative, this structural complexity hinders their ability to manage codebases directly. We introduce the \underline{\textbf{P}}rovable \underline{\textbf{R}}epresentation \underline{\textbf{O}}f \underline{\textbf{O}}riginal \underline{\textbf{F}}unctionality (\modelbf{}), which manages codebases indirectly via structured specifications. To enable full-lifecycle codebase management strictly through these specifications, \model{} abstracts codebase topology into a hierarchical natural-language representation. To establish absolute trust, the system proves semantic equivalence by reconstructing source code exclusively from this specification. This verified foundation drives maintenance requests, executing code modifications while synchronously updating itself to prevent semantic drift. Experiments on real-world repositories confirm the effectiveness of these specifications.
\end{abstract}

\section{Introduction}

\begin{figure}[t]
  \centering
  \includegraphics[width=\columnwidth]{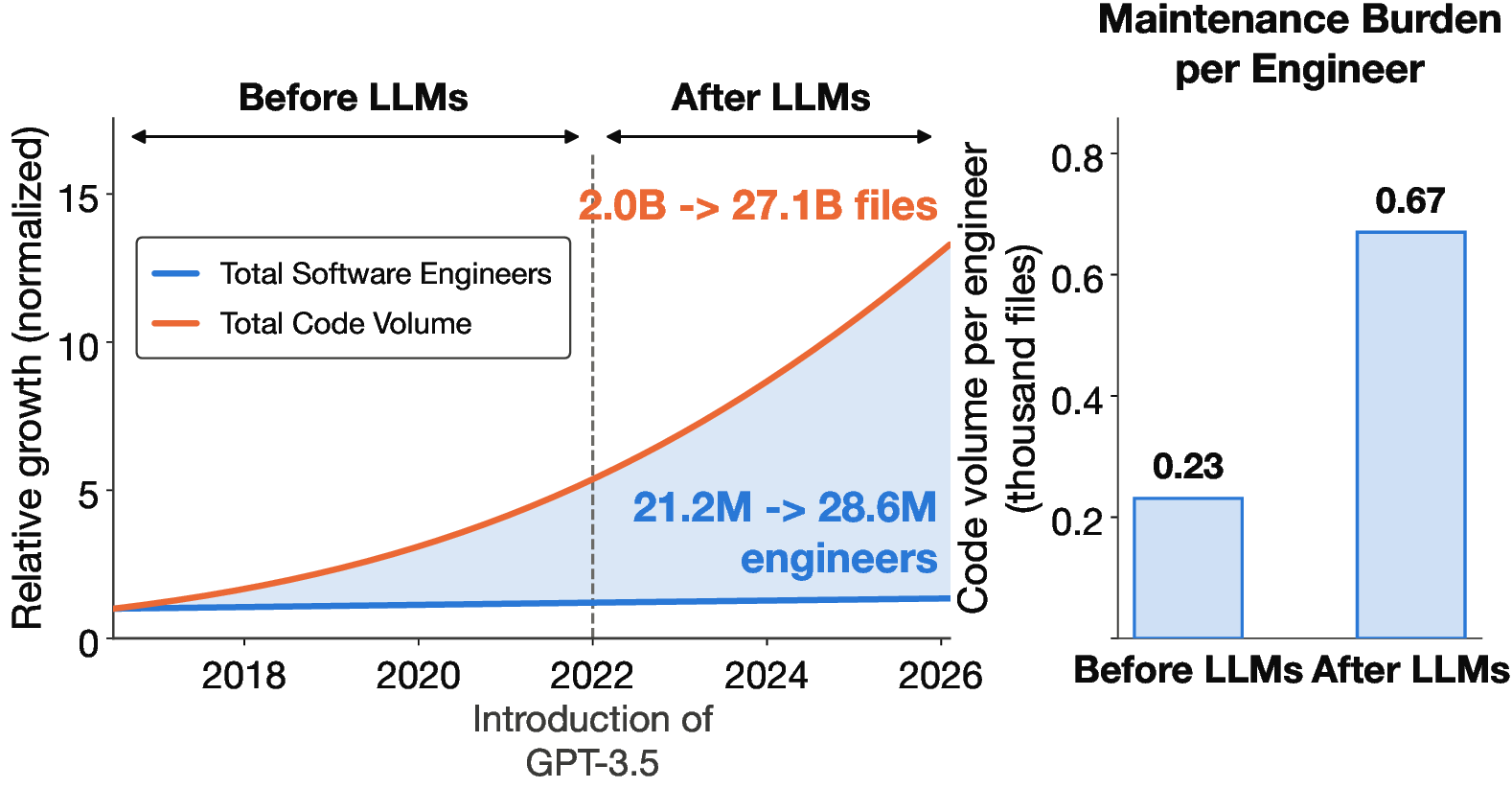}
  \caption{Trends in total code volume versus the software engineering workforce, and the corresponding average maintenance burden per engineer (2016–2026).}
  \label{fig:intro-gap}
\end{figure}

Software systems are experiencing continuous growth in scale and structural complexity, a trend accelerated by the proliferation of Large Language Models (LLMs)~\cite{hatton2017long,ouyang2025repograph}. Consequently, code generation rates outpace the cognitive capacities of engineering teams. As Figure~\ref{fig:intro-gap} illustrates, this widening disparity elevates the maintenance burden per engineer, resulting in mounting technical debt as machine-generated code is integrated without adequate human review or long-term maintenance.

A potential approach to mitigating this maintenance burden is to utilize LLMs not only as generators but also as automated maintainers~\cite{liu2024repobench}. However, applying LLMs to repository-level code management is impeded by two bottlenecks: context window constraints, which preclude processing repositories simultaneously~\cite{liu2024lost}, and global dependencies that span multiple files~\cite{jimenez2024swe}. Together, these limitations induce context degradation, leading to code hallucination and logical inconsistencies.

To bypass these direct-processing limitations, a viable strategy is to manage repositories indirectly through an intermediate representation (IR)~\cite{lattner2004llvm,iyer2016summarizing}, allowing models to reason over structural abstractions rather than raw, unbounded source code~\cite{hejderup2018software}. To ensure reliability, this IR should satisfy three properties: (1) granularity adaptation, partitioning the repository into localized chunks that fit within context limits; (2) topology preservation, retaining the global dependency structures; and (3) verifiability, ensuring the representation is provably equivalent to the original code behavior rather than an unverified generative approximation.

To realize this paradigm, we apply the standard software specification (\textbf{Spec}) as a structured intermediate representation. At the micro-level, it encapsulates objects and functions as minimal closure units, utilizing modular natural language to capture essential semantics. At the macro-level, these units are structurally linked via the repository's native call graph. This dual-level architecture adapts to LLM context limits while preserving global topology, establishing an explicit, human-readable abstraction layer that facilitates code comprehension and editing without requiring developers to navigate the underlying syntax directly.

Furthermore, the effectiveness of this representation depends entirely on its rigorous verifiability. To prove that a Spec objectively reflects the source code rather than LLM hallucinations, we introduce a strict round-trip equivalence verification framework. By extracting a Spec from the original code and reconstructing the code relying exclusively on this Spec without source reference, we can objectively validate that the representation captures all critical functionality without omission, rendering it trustworthy for downstream operations.

We decompose the realization of this representation into three lifecycle questions. \textit{Generation}: how can this structured Spec be completely and systematically constructed for large-scale repositories? \textit{Verification}: how can its precise consistency with the original codebase be objectively validated to prevent generative hallucinations? \textit{Update}: how can a verified Spec drive code modifications, and how can bi-directional synchronization be maintained across sustained maintenance cycles to prevent semantic drift?

To systematically address these questions, we developed a closed-loop system, the \underline{\textbf{P}}rovable \underline{\textbf{R}}epresentation \underline{\textbf{O}}f \underline{\textbf{O}}riginal \underline{\textbf{F}}unctionality (\modelbf{}), which enables developers to manage code strictly by managing Spec.

\paragraph{Contributions.} Our main contributions include:
\begin{itemize}[leftmargin=0.12in]
    \item A systematic methodology for Spec generation: a hierarchical natural-language representation mirroring the repository's structural organization, driven by an algorithm that enables LLMs to process the codebase comprehensively.

    \item A rigorous validation framework for the generated Spec: evaluating internal quality and semantic consistency with the source code, establishing it as an objectively verifiable equivalent rather than an unverified generative assumption.

    \item A methodology for Spec-driven codebase updates: \model{} modifies and validates code while synchronously updating the Spec, enabling developers to manage code exclusively through Spec without semantic drift.
\end{itemize}

\section{Related Work}
\label{sec:related-work}
Large codebases increase the burden on LLMs attempting to modify code directly, elevating the risk of broken dependencies and context-window overflow. To mitigate these bottlenecks, IR are constructed to indirectly constrain codebase updates. The evolution of these representations spans natural language abstraction, graph-based structural modeling, and round-trip verification, establishing the theoretical foundation for our methodology.

\subsection{Natural Language Representations}

Code summarization was framed as sequence transduction, translating code into natural language descriptions~\cite{iyer2016summarizing}. Bimodal pre-training introduced identifier-aware objectives for bidirectional language-code mapping~\cite{wang2021codet5}. This extended to repository-level evolution: natural language outlines synchronized with code~\cite{shi2025natural}, RepoAgent's incremental updates from call relations~\cite{luo2024repoagent}, and DocAgent's multi-agent, dependency-order generation~\cite{yang2025docagent} drive changes through natural language. These methods read well, but free text lacks structural constraints and verifiable semantics, producing ambiguity on complex logic and limiting safe modification.

\subsection{Graph-Based Structural Models}

Graph structures represent code dependencies, data flow, and call logic: variable-level data-flow graphs capture a function's internal logic during pre-training~\cite{guo2020graphcodebert}, repository-level definition-reference graphs support issue localization and repair~\cite{ouyang2025repograph}, a repository planning graph guides step-by-step module generation~\cite{luo2026rpg}, RPG-Encoder~\cite{luo2026closing} and LocAgent~\cite{chen2025locagent} reverse-model codebases into a persistent graph for representation and localization, and CoSIL~\cite{jiang2025issue} constructs call graphs during search. These methods preserve macro-level dependencies but encode nodes as high-level intent or vectors, lacking detail for fine-grained modification.

\subsection{Round-Trip Verification}

Verifying that IR-space modifications map correctly to executable code has become an independent direction, round-trip consistency: RTCE~\cite{maveli2026can} formalizes it as a benchmark on encoding-decoding fidelity, SimRAG~\cite{xu2025simrag} applies conformal prediction to filter inconsistent samples, and EPAM~\cite{grynets2026specification} drives industrial code migration through a IR-mediated code-text-code architecture. These pipelines remain incomplete: verification stays soft, relying on structural similarity or self-assessment, with no continuous drift detection.

\section{Problem Formulation}
\label{sec:problem-formulation}

A codebase update modifies $C$, with test suite $T$, according to change request $R$ to produce codebase $C'$ while preserving quality. Its logic parses into an abstract syntax tree $\mathrm{AST}(C)$; $C$ exhibits a nested hierarchy of levels $\mathcal{L}$ (modules decomposing into function clusters), formalized as a directed graph $D_C = (N_C, E_C)$ whose nodes distribute across $\mathcal{L}$ and edges reflect containment and dependency.

A safe update requires $C'$ to satisfy a dual equivalence constraint, $\mathcal{E}(C' \mid C, R) = 1$: structural equivalence $\mathrm{EQUIV}_{struct}(C, C')$, requiring unaffected regions' syntactic framework to remain unchanged, $\mathrm{AST}(C \setminus R) \cong \mathrm{AST}(C' \setminus R)$; and behavioral equivalence $\mathrm{EQUIV}_{beh}(C', T)$, requiring identical outputs over the unchanged domain, $\forall x \in \mathcal{X} \setminus R, C'(x) = C(x)$.

Existing methods solve mapping $\mathcal{F}: (C, R) \to C'$ directly in code space $\mathcal{C}$, tracking dependency structure while handling syntactic detail. As scale grows and LLM-generated code enters in bulk, information density in $D_C$ rises sharply, rendering direct solving of $\mathcal{F}$ fragile and prone to breaking $\mathcal{E}$ through cascading modifications.

To resolve this bottleneck, an IR consistent with the code's topological structure but easier to reason about semantically can overcome the limits of manipulating low-level code directly. We formalize this representation's set as a surrogate space $\mathcal{S}$, recasting codebase updates as a structure-preserving semantic distillation across spaces.

Since $D_C$'s topological dependencies determine system stability, we define a mapping $\Phi: \mathcal{C} \to \mathcal{S}$ from the code space $\mathcal{C}$ to a surrogate space $\mathcal{S}$, keeping the graph structure unchanged. An initial code $C$ is instantiated as a Spec graph $S = \Phi(C)$: $\Phi$ preserves the isomorphic relation $D_C \cong D_S$, retaining $\mathcal{L}$ and $E_C$, while filtering out language-specific detail at the node level, mapping code to natural-language Spec nodes $N_S$. The codebase update then becomes:
\begin{equation}
    C' = \Phi^{-1}\Big(\Delta(S, R)\Big),
    \label{eq:code-evolution}
\end{equation}
where $S = \Phi(C)$ and $\mathcal{E}(C' \mid C, R) = 1$, omitting syntactic detail while holding topological complexity constant, retaining only essential code intent as a semantic representation humans can manipulate directly. We instantiate $S$ as a natural-language Spec, detailed in the next section.

\section{Methodology}
\label{sec:methodology}

\begin{figure*}[t]
  \centering
  \includegraphics[width=\textwidth]{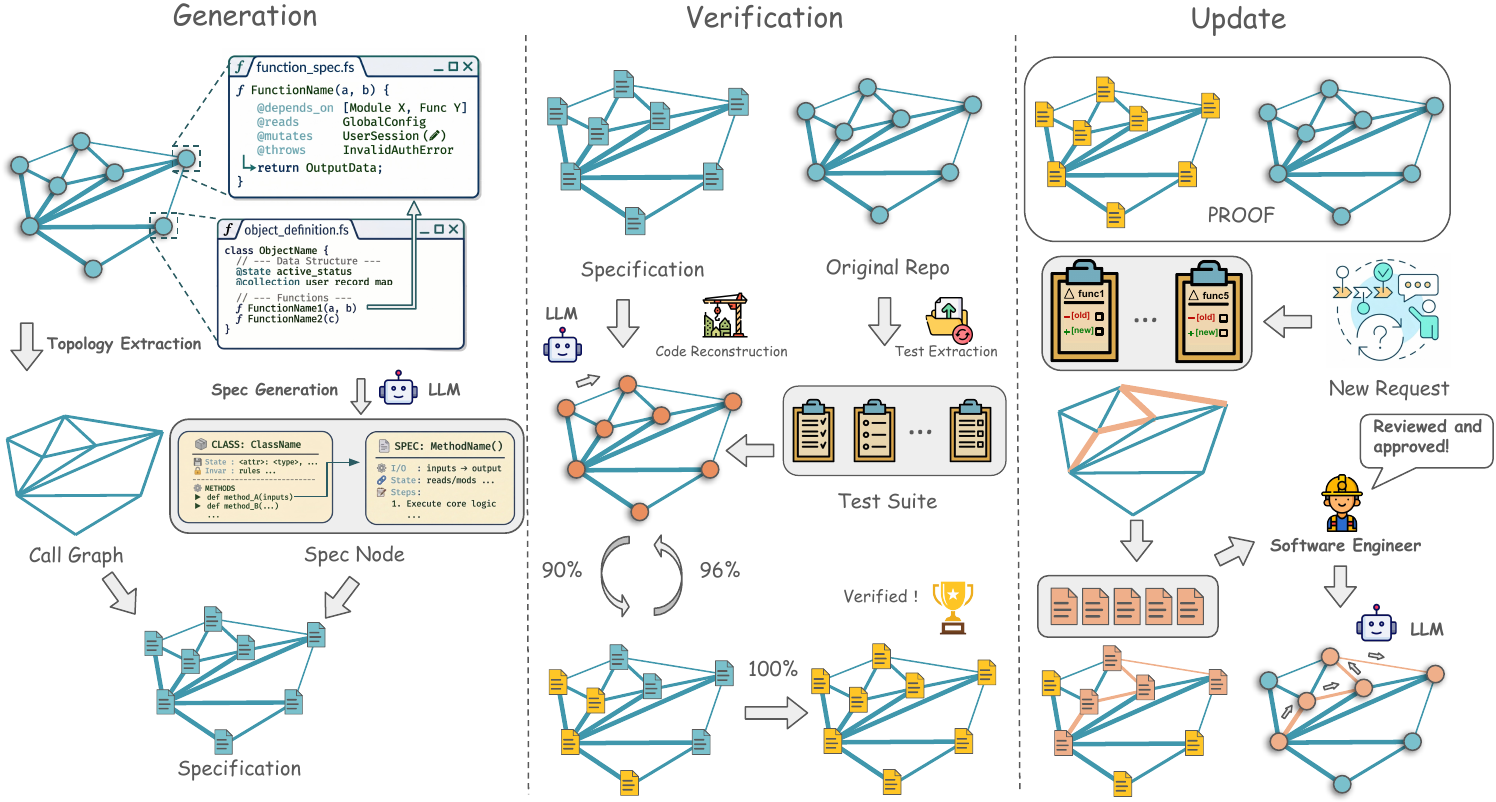}
  \caption{Overview of \model{}'s three-phase pipeline. \textbf{Generation}: the codebase's call graph is traversed node by node, and every Object and Function receives a structured natural-language Spec node, forming the Spec, an interconnected graph of Spec nodes mirroring the exact shape of the code's dependency graph. \textbf{Verification}: code is reconstructed from the Spec alone and checked against the original codebase's own test suite over repair rounds, synchronously updating both the code and the Spec until every originally-passing test passes on the reconstruction, thereby validating the Spec. \textbf{Update}: given a new change request, \model{} localizes the affected Spec nodes, the natural-language edit is reviewed and approved by a software engineer, and the approved edit is propagated into code by the LLM.}
  \label{fig:pipeline}
\end{figure*}

To address direct code management limitations, we propose an end-to-end approach called \model{}. Built around the Spec formalized in Section~\ref{sec:problem-formulation}, it enables safe codebase updates by manipulating natural language Spec. Figure~\ref{fig:pipeline} shows three pipeline phases: Generation, Verification, and Update.

\subsection{Specification Generation}
\label{sec:hierarchy-design}

\begin{figure}[t]
  \centering
  \includegraphics[width=\columnwidth]{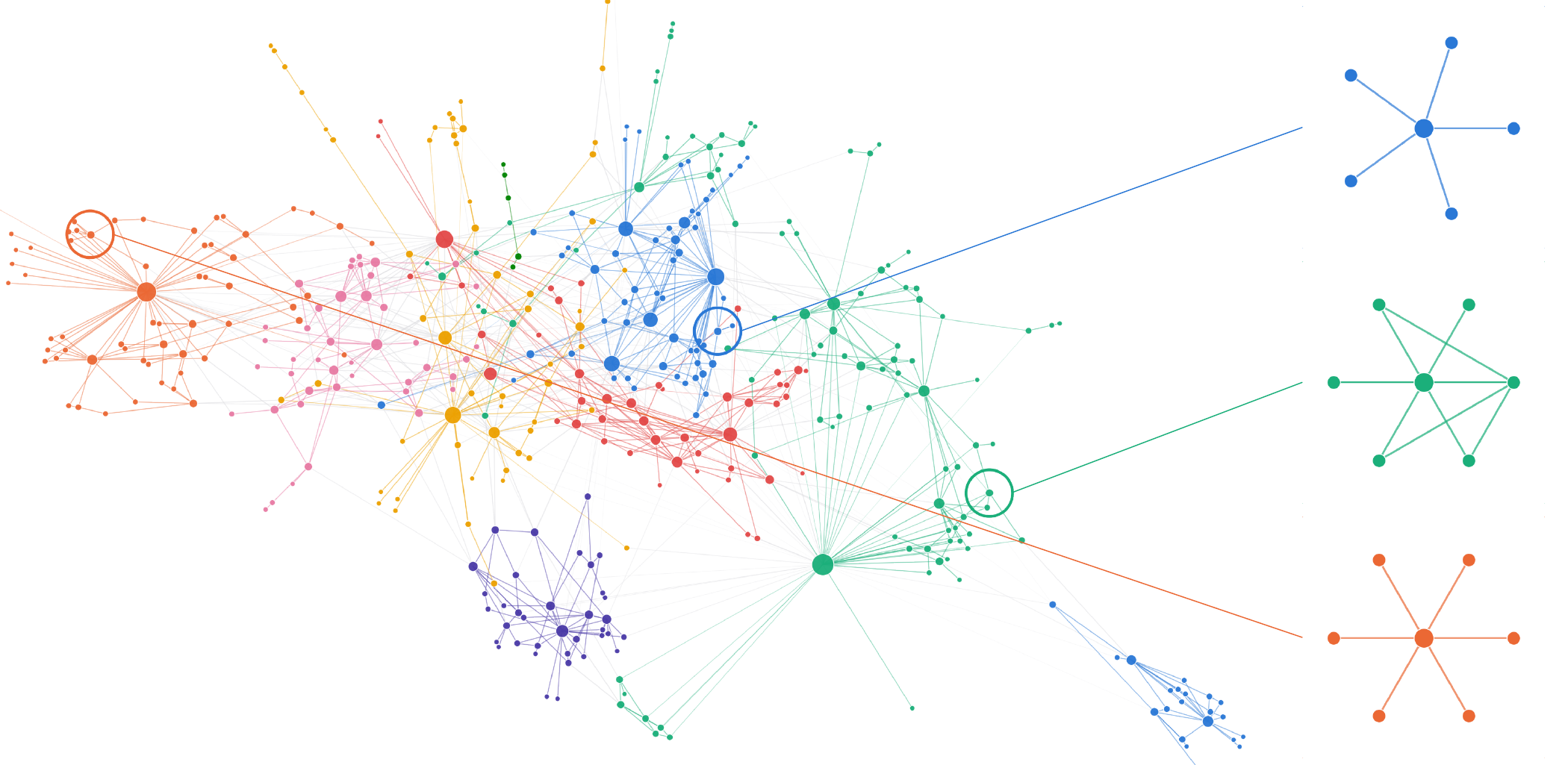}
  \caption{Function-level call graph of Flask's core module (436 nodes, 1016 edges).}
  \label{fig:callgraph}
\end{figure}

Determining the update logic directly in the low-level code space faces high complexity: Figure~\ref{fig:callgraph} shows the call graph of Flask (a real-world codebase), whose structure $D_C = (N_C, E_C)$ forms a dense, entangled network at the macroscopic level, making full context difficult to process directly for developers and LLMs alike, obstructing the mapping $\Phi$ into the surrogate space.

Although the macroscopic network is densely entangled, Figure~\ref{fig:callgraph}'s right panel shows that microscopic dependencies are highly localized: any given node's local dependencies are confined to a few immediate components.

Decomposing the mapping at the local dimension is thus more viable than tackling dependencies head-on: the system distills local code fragments per node, then scales these up by preserving the call edges $E_C$, assembling a global Spec graph $S = (N_S, E_S)$ isomorphic to $D_C = (N_C, E_C)$, $S \cong D_C$, where $N_S$ is the distilled Spec modules and $E_S$ their dependency links.

The following question is how to define this decomposition's granularity: to preserve execution logic completely, we partition the space along the codebase's static nested hierarchy $\mathcal{L}$, preserving original semantic boundaries and dependencies.

We instantiate $\mathcal{L}$ top-down into five levels, each with a physical form and semantic function:

\begin{itemize}[leftmargin=0.2in, itemsep=0pt, parsep=0pt, topsep=3pt]
  \item \textbf{Project}: the top-level codebase abstraction.
  \item \textbf{Module}: a physical directory or file grouping classes and functions by business domain.
  \item \textbf{Object}: the union of a data structure (state) and Functions (behavior), its Spec aggregated from its functions.
  \item \textbf{Function}: smallest, self-contained execution unit with defined inputs, outputs, and logic, serving as $\Phi$'s semantic distillation entry point.
  \item \textbf{Segment}: the code lines forming a single Function (a loop, or branch), needed since functions are often too long for a model to digest at once and must be split for processing.
\end{itemize}

This static decomposition, however, severs the dynamic call edges $E_C$. To restore the code topologically, we design the core mapping nodes, Object and Function, as \emph{semantic closures} with strict constraints and self-contained boundaries.

Existing LLM-based summarization methods mostly use free-text descriptions, readable but blurring precise dependencies and making $E_C$ harder to track. To maintain topological isomorphism in $\mathcal{S}$, we constrain the model's free generation through a standardized, strictly constrained template.

The basic carrier of the behavioral contract is the \textbf{Function closure}: every function node $v_{func} \in N_S$ is a structured tuple containing its segmented control-flow pieces (Segments), sealing within its boundary the external logic and side effects it depends on, following a six-field contract, $v_{func} = \langle \text{Desc}, \text{I/O}, \text{Meta}, \text{Deps}, \text{Data}, \text{Error} \rangle$:
\begin{itemize}[leftmargin=0.2in, itemsep=0pt, parsep=0pt, topsep=3pt]
  \item \textbf{Description}: the execution intent;
  \item \textbf{I/O Definition}: constraints on parameter types and return values;
  \item \textbf{Meta}: programming language, purity (side effects), and atomicity (multi-threaded safety);
  \item \textbf{Dependencies \& External Calls}: every call to another node;
  \item \textbf{Data \& Mutations}: which global state or mutable arguments are read or modified;
  \item \textbf{Error Handling}: the boundaries of exceptions thrown and caught.
\end{itemize}

Through the \texttt{Deps} and \texttt{Data} fields, the system records every function's interaction with its external environment, specifically detailing which nodes it calls and which global state it reads or modifies, thereby converting dependencies previously hidden in the code into static, structured declarations.

In the object-oriented paradigm, a class encapsulates its variables and data structures, so we define the \textbf{Object closure} as a nested tuple $v_{obj} \in N_S$: a class-level state definition enclosing methods $M_{obj}$ operating on that state, $v_{obj} = \langle \text{ClassDesc}, \text{DataState}, \text{Invariants}, M_{obj} \rangle$. The three outer fields are:
\begin{itemize}[leftmargin=0.2in, itemsep=0pt, parsep=0pt, topsep=3pt]
  \item \textbf{ClassDesc}: the object's design intent and responsibility;
  \item \textbf{DataState}: the class's internal instance variables and their type constraints;
  \item \textbf{Invariants}: the logical constraints that must hold throughout an instance's lifecycle.
\end{itemize}
An object's internal behavior aggregates from $M_{obj}$: each method $v_m \in M_{obj}$ reuses the function tuple structure above ($v_m = \langle \text{Desc}, \text{I/O}, \dots \rangle$), except its Data field extends to and is constrained by the parent's DataState (the class's \texttt{self} variable).

These definitions provide the foundation for scaling through the hierarchy: since every node encapsulates both its internal logic and an interface declaration, discrete code entities are stitched back together via these cues, aggregating locally into Modules and composing upward into a network reflecting the project's global dependency relations.

Relying on these interface declarations, the system splits $\Phi$ into two phases: local node instantiation, and global topology extraction and assembly.

In the local instantiation phase, the system traverses the codebase along its static hierarchy; for each node the LLM reads its code, splits it into ordered control-flow segments, and fills the six contract fields per the template, keeping context small and reducing drift from long-range dependencies.

In the global assembly phase, the system reconstructs the dependency network among nodes: since every closure declares its call intentions through the \texttt{Dependencies \& External Calls} field, a deterministic parsing algorithm reads these fields and extracts corresponding edges, without the LLM inferring the macroscopic structure.

This avoids the structural bias free-form connection generation could introduce, aggregating related nodes locally into Modules and composing them upward, scaling from the smallest nodes to a global specification graph $S = \Phi(C)$ that remains topologically isomorphic to $C$.

\subsection{Equivalence Verification}
\label{sec:mapping-validation}

The previous section defined the mapping $\Phi$ from $\mathcal{C}$ to $\mathcal{S}$. During generation, structured closures and deterministic parsing guarantee that the resulting Spec graph $S = \Phi(C)$ is isomorphic in topology to the original codebase, $D_C \cong D_S$.

Topological isomorphism only guarantees correct calls and dependencies between nodes. Each node's natural-language description is written by the language model, and if it omits or misdescribes information while carrying out $\Phi$, $S$ will not accurately reflect the behavior of $C$ even when the structure is correct.

Structural isomorphism is therefore not sufficient on its own. Since $S$ is the only object later localization and editing act on, we must show that $\Phi$ preserves the information needed to reproduce the code's behavior, capturing everything that determines how the code behaves when run rather than mere syntactic details or formatting. Only once this is checked, confirming $S$ accurately represents $C$'s behavior, are the later editing procedures justified.

$S$ and executable code $C$ are different representations that cannot be compared directly. To check whether $S$ contains enough behavioral information, we use a zero-edit roundtrip: reconstruct code from $S$ via the inverse mapping, $\hat{C} = \Phi^{-1}(S)$, turning whether $S$ is correct into whether $C$ and $\hat{C}$ behave the same way. We write $\hat{C}$ to distinguish this from $C'$, reserved for code produced by an actual change request $R$ in Section~\ref{sec:change-localization}.

Two pieces of code are behaviorally equivalent if they produce the same output on every input, matching the definition of $\mathrm{EQUIV}_{beh}$ in Section~\ref{sec:problem-formulation}: in the zero-edit case ($R = \emptyset$), $\hat{C}$ should produce the same output as $C$ for every $x \in \mathcal{X}$:
\begin{equation}
  \forall x \in \mathcal{X}, \; \hat{C}(x) = C(x)
\end{equation}
If so, $S$ preserves behavior-determining information in $C$, and $\Phi$ loses no information that matters.

Checking this equality requires examining every input in $\mathcal{X}$, which is not possible in practice; we therefore use the repository's existing test suite $T$ as a finite sample of $\mathcal{X}$ to approximate it.

Requiring $\hat{C}$ to pass every test in $T$ runs into a practical problem: real code $C$ often has existing defects or flaky tests and does not itself pass every test in $T$, so this would require the model to fix pre-existing defects during reconstruction, which this check is not meant to verify.

To resolve this, we relax the equivalence condition. Let $\mathrm{Pass}(C) = \{t \in T : \mathrm{exec}(C, t) = \text{pass}\}$ be the tests $C$ passes, using the same $\mathrm{exec}$ from Section~\ref{sec:problem-formulation}; we replace exact equivalence with directed inclusion:
\begin{equation}
  \mathrm{Pass}(C) \subseteq \mathrm{Pass}(\hat{C})
\end{equation}
This states the minimum requirement, no regression: it does not penalize $\hat{C}$ for defects already in $C$, verifying only that logic $C$ already implements correctly is not lost after the roundtrip.

This set-inclusion condition is binary; to quantify and compare it across experiments, we turn it into a continuous score, \textbf{Fidelity}: the fraction of $C$'s originally passing tests that $\hat{C}$ still passes.
\begin{equation}
  \mathrm{Fidelity}(N_S) = \frac{|\mathrm{Pass}(C) \cap \mathrm{Pass}(\hat{C})|}{|\mathrm{Pass}(C)|}
  \label{eq:fidelity}
\end{equation}
where $\hat{C} = \Phi^{-1}(N_S)$. If $C$ already passes the entire test suite ($\mathrm{Pass}(C) = T$), the denominator becomes $|T|$, and Fidelity reduces to the ordinary test pass rate.

This continuous score avoids treating the mapping as a failure over a few failing tests while keeping the check strict: a high score, whether reported as Fidelity on real code with pre-existing defects or as the pass rate on a clean benchmark, is evidence that $S$ preserves the behavior-determining information in $C$. Passing this check is the precondition for the localization and editing procedures described in the following sections.

\subsection{Codebase Update}
\label{sec:change-localization}

When the system receives a change request $R$, the large language model maps it onto the specification graph $S$ in the surrogate space $\mathcal{S}$ and locates the affected subset of nodes $N_R$. Through this process, the global modification operation $\Delta(S, R)$ is decomposed into a set of mutually independent local changes (i.e., an independent modification $\Delta v$ is performed for each affected node $v \in N_R$), meaning the macro-level requirement is concretely translated into separate updates to specific text fields within a function or object tuple. Because the number of affected nodes is far smaller than the total number of nodes in the Spec graph (i.e., $\vert N_R \vert \ll \vert N_S \vert$), this design of decomposing the request into node-level operations mechanically confines the semantic error a single change may introduce to a local region.

At the same time, when the system applies the inverse mapping $C' = \Phi^{-1}\big(\Delta(S, R)\big)$ to reconstruct the code, because the modification is confined to specific nodes, the syntactic structure of unaffected regions remains unchanged. This satisfies the equivalence constraint defined earlier, $\mathrm{AST}(C \setminus R) \cong \mathrm{AST}(C' \setminus R)$. This mechanism not only narrows the scope of subsequent testing and verification but also effectively limits the system state's drift within a single change.

In this process, the focus of code review and editing shifts to the semantic level. Because the modification proposal for the affected nodes $N_R$ is presented as structured natural language, a developer need only read the text and confirm whether its business logic matches the change request. Once the modification proposal has passed review within the surrogate space, the generation and assembly of the underlying code (i.e., applying the inverse mapping $\Phi^{-1}$) is carried out automatically by the large language model. This mechanism establishes a division of labor in which the developer manages the Spec and the model modifies the code, so that the developer does not need to directly handle low-level syntactic detail.

Although local mapping can effectively control the error of a single modification, after $i$ successive incremental updates, some semantic drift between the current Spec graph $N_S^{(i)}$ and the underlying code remains difficult to avoid.

To measure this drift, the system provides an intuitive calibration method. When necessary, the system can regenerate a fully accurate reference Spec graph $N_S^{(i)*}$ from the current underlying code, and quantify the error by computing the difference in Fidelity between the two:
\begin{equation}
  \Delta_i = \mathrm{Fidelity}\big(N_S^{(i)*}\big) - \mathrm{Fidelity}\big(N_S^{(i)}\big)
\end{equation}
Here, $\Delta_i$ serves only as a reference indicator; the system does not force any reset based on it. Depending on the situation, when a developer judges the drift to be too large, they may choose to overwrite the current state with the reference graph, thereby realigning the code and the Spec. This design both preserves the flexibility of day-to-day updates and provides a means of removing long-term accumulated error, ensuring the codebase remains stable and reliable over long-term evolution.

Appendix~\ref{sec:appendix-spec-example} walks through a concrete case study of how a Spec node is structured in practice and how a change request $R$ is decomposed into node-level edits kept in sync with code.

\section{Experiments}
\label{sec:experiments}

\subsection{Experimental Setup}
\label{sec:exp-setup}

We evaluate on three SWE-bench~\cite{jimenez2024swe} repositories (9.5K--37K lines, 0.49K--4.2K tests; Table~\ref{tab:repo-scale}). Original test suites are isolated during generation to prevent leakage. In our primary experiments, we apply Claude 5 Sonnet as the underlying language model. We provide a detailed ablation study regarding token efficiency and backbone sensitivity in Appendix~\ref{sec:appendix-ablation-tokens}.

\begin{table}[h]
\centering
\small
\begin{tabular}{lrrrr}
\toprule
Repo & Files & Lines & Nonblank & Tests \\
\midrule
Flask & 24 & 9{,}502 & 7{,}560 & 491 \\
Seaborn & 54 & 29{,}224 & 24{,}154 & 2{,}381 \\
Pytest & 81 & 37{,}770 & 31{,}463 & 4{,}224 \\
\bottomrule
\end{tabular}
\caption{Scale of the three evaluation repositories.}
\label{tab:repo-scale}
\end{table}

We compare against three baselines (Section~\ref{sec:related-work}): \textbf{RepoAgent}~\cite{luo2024repoagent}, \textbf{RPG-Encoder}~\cite{luo2026closing}, and \textbf{EPAM}~\cite{grynets2026specification}. Because comparing heterogeneous IRs directly is difficult, we assess information retention by reconstructing codebases strictly from IR (without source access) across three dimensions:

\textbf{Behavioral equivalence}: Fidelity (Equation~\ref{eq:fidelity}) measures the reconstruction's pass rate, calibrated against the original suite's perfect pass rate. Pass@1 is the initial reconstruction's pass rate. Iterative failure-driven repairs yield Pass@k, strictly maintaining test suite isolation. In our experiments, we set $k=3$, recording the progressive success rates from Pass@1 through Pass@3.

\textbf{Structural equivalence}: We compute Abstract Syntax Tree (AST) similarity between the original codebase and the unmodified Pass@1 reconstruction using \texttt{pycode\_similar}~\cite{fyrestone2017pycodesimilar}.

\textbf{Interpretability}: We average scores across four dimensions evaluated by 3 independent human raters and 3 LLM-as-a-judge trials.

\subsection{Behavioral Equivalence}
\label{sec:exp-roundtrip}

\begin{table}[h]
\centering
\small
\begin{adjustbox}{max width=\columnwidth}
\begin{tabular}{llrrr}
\toprule
Repo & Method & Pass@1 & Pass@2 & Pass@3 \\
\midrule
\multirow{4}{*}{Flask} & \model{} & \textbf{84.9\%} & \textbf{97.6\%} & \textbf{100\%} \\
 & EPAM & 31.77\% & 94.30\% & 98.57\% \\
 & RepoAgent & 4.28\% & 67.5\% & 94.6\% \\
 & RPG-Encoder & 57.8\% & 94.0\% & 99.2\% \\
\addlinespace
\multirow{4}{*}{Seaborn} & \model{} & \textbf{83.5\%} & \textbf{98.1\%} & \textbf{100\%} \\
 & EPAM & 0.00\% & 35.78\% & 59.64\% \\
 & RepoAgent & 29.95\% & 46.96\% & 51.32\% \\
 & RPG-Encoder & 3.49\% & 26.84\% & 38.93\% \\
\addlinespace
\multirow{4}{*}{Pytest} & \model{} & 0\% & \textbf{91.24\%} & \textbf{96.80\%} \\
 & EPAM & 0.00\% & 0.00\% & 8.07\% \\
 & RepoAgent & 0.09\% & 74.48\% & 89.75\% \\
 & RPG-Encoder & 0.38\% & 86.60\% & 90.46\% \\
\bottomrule
\end{tabular}
\end{adjustbox}
\caption{Test-suite pass rate across repair rounds.}
\label{tab:roundtrip-results}
\end{table}

\begin{figure}[t]
  \centering
  \includegraphics[width=\columnwidth]{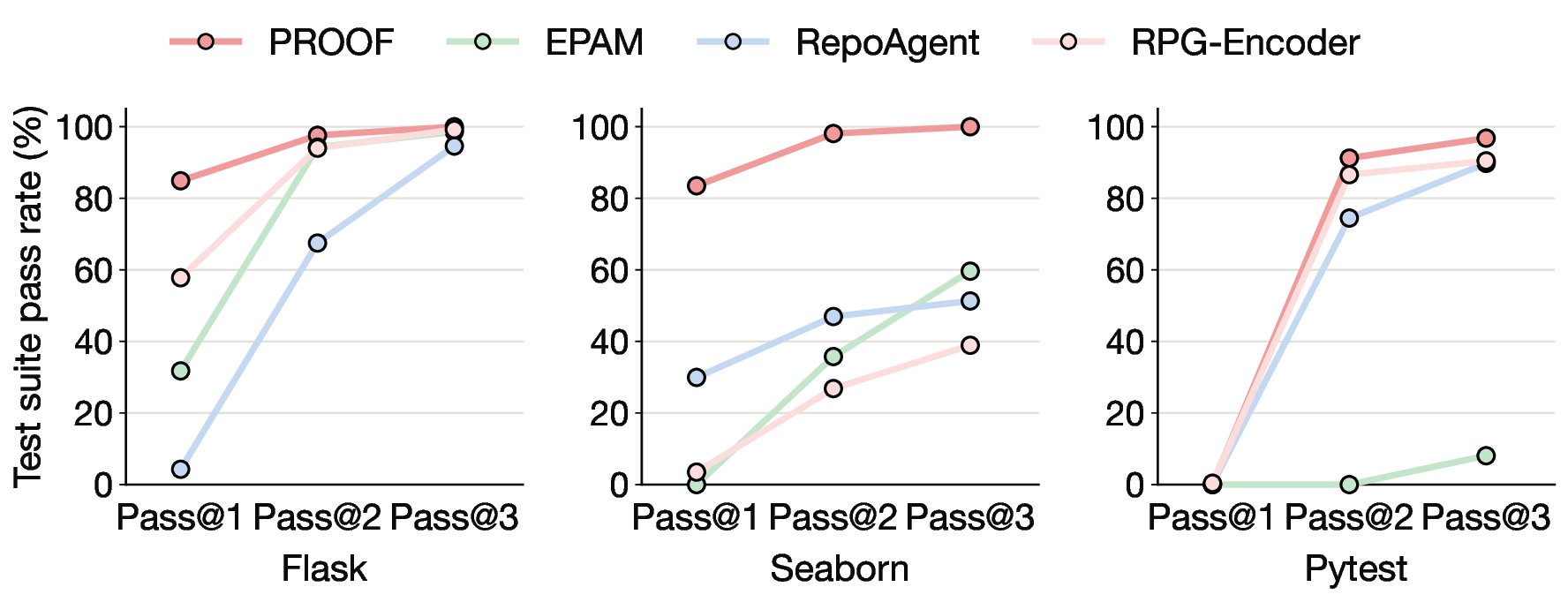}
  \caption{Test-suite pass rate across repair rounds (Pass@1 $\to$ Pass@2 $\to$ Pass@3).}
  \label{fig:roundtrip-passk}
\end{figure}

As shown in Table~\ref{tab:roundtrip-results}, \model{} yields the highest Pass@3 across all three repositories (100\%, 100\%, and 96.80\%). For Flask and Seaborn, \model{} demonstrates higher Pass@1 and reaches 100\% within three rounds. The variance among methods is most prominent on Seaborn: \model{} reaches 100\% at Pass@3, whereas RepoAgent and RPG-Encoder record 51.32\% and 38.93\%, respectively.

The results on Pytest warrant further analysis. As the largest repository, Pytest exhibits the highest code volume, test count, and structural coupling. All evaluated methods record a Pass@1 of approximately 0\%. This occurs because high-level integration errors in a tightly coupled architecture propagate extensively, leading to widespread test failures rather than indicating that all reconstructed functions are incorrect. However, by preserving call graph topologies and explicit node boundaries, \model{} facilitates structural error localization during the repair phase, bringing the pass rate to 91.24\% at Pass@2. Similarly, RPG-Encoder, which also incorporates graph structures, improves to 86.60\% at Pass@2. In contrast, EPAM, which operates without strict boundary definitions, records a 0.00\% pass rate at Pass@2.

\subsection{Structural Equivalence}
\label{sec:exp-similarity}

\begin{table}[h]
\centering
\small
\begin{adjustbox}{max width=\columnwidth}
\begin{tabular}{lrrr}
\toprule
Method & Flask & Seaborn & Pytest \\
\midrule
\model{} & \textbf{90.45\%} & \textbf{87.32\%} & \textbf{96.12\%} \\
EPAM & 83.22\% & 71.91\% & 80.90\% \\
RepoAgent & 82.43\% & 80.25\% & 86.52\% \\
RPG-Encoder & 89.67\% & 40.36\% & 86.10\% \\
\bottomrule
\end{tabular}
\end{adjustbox}
\caption{Structural similarity to the original codebase.}
\label{tab:similarity-results}
\end{table}

\begin{figure}[t]
  \centering
  \includegraphics[width=\columnwidth]{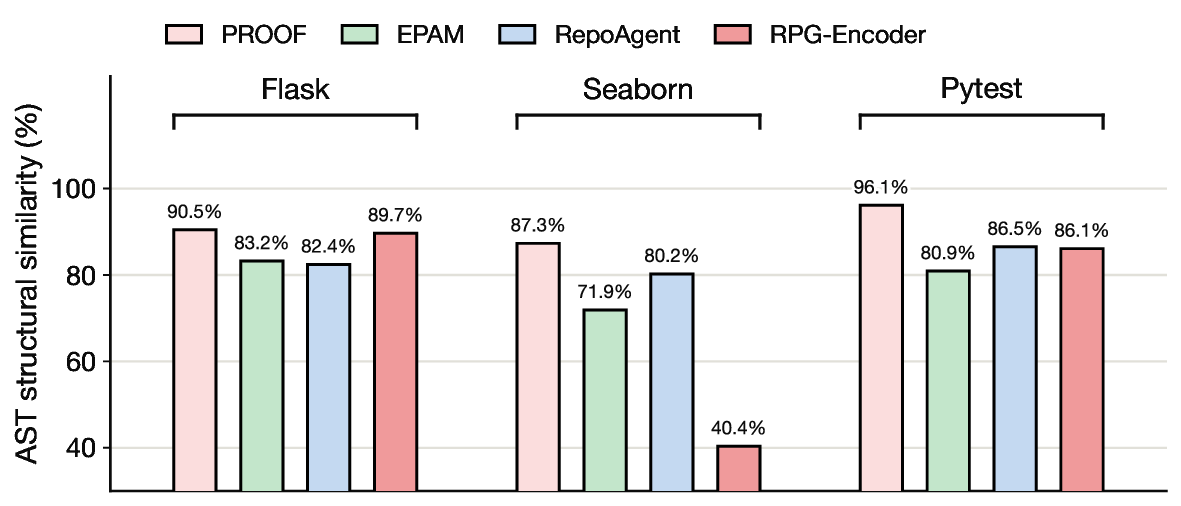}
  \caption{AST structural similarity (\texttt{pycode\_similar}) between each system's reconstructed/regenerated codebase and the real source.}
  \label{fig:similarity-comparison}
\end{figure}

To evaluate architectural preservation, we measure Abstract Syntax Tree (AST) similarity between reconstructed and ground-truth codebases. As Table~\ref{tab:similarity-results} shows, \model{} achieves the highest similarity globally, peaking at 96.12\% on Pytest. This highlights the necessity of combining macro-level topology with fine-grained closures. While RPG-Encoder performs competitively on Flask (89.67\%), its abstraction of node intents causes structural collapse on Seaborn (40.36\%). Conversely, \model{} maintains 87.32\% fidelity by explicitly defining boundaries for every node.

Baselines also reveal the limitations of pure natural language and soft verification. Lacking rigid constraints, RepoAgent suffers generative deviations capping at 86.52\%. Similarly, EPAM's soft verification permits continuous drift, yielding 80.90\% on Pytest. \model{} resolves these issues by enforcing the Spec as a strict constraint. By isolating functions into verifiable closures and utilizing continuous validation, our approach eliminates ambiguity and ensures precise AST alignment.

\subsection{Interpretability}
\label{sec:exp-trust}

We sample 7 functions each from Flask, Seaborn, and Pytest (21 total), including 2 dependency-linked groups per repository so Consistency has real cross-node context to check. Three raters independently score the four anonymized descriptions (\model{} and baselines EPAM, RepoAgent, RPG-Encoder) on four dimensions: Faithfulness (match to code's real behavior), Verifiability (specific, checkable claims), Completeness (I/O, side effects, error conditions, dependencies covered), and Consistency (agreement with parent/sibling nodes). Together, these criteria provide a sufficient framework to ensure both local node accuracy and global structural coherence, verifying the representation is rigorous enough to safely drive automated modifications. An LLM-as-a-Judge scores the same items with the identical rubric, repeated three times per item; we take the mean, with run-to-run variance as a stability check.

\model{} ranks first on all four dimensions under both scoring sources, though the margin varies. Faithfulness and Verifiability gaps are small (0.3 and 0.5 points), since even free-text baselines mostly state a function's behavior clearly. Completeness and Consistency gaps are larger: on Completeness, \model{} exceeds RepoAgent (3.3) by 1.1 points and EPAM (1.9) by 2.5, tracking to the six-field closure template that forces every node to state I/O, side effects, error conditions, and dependencies explicitly. On Consistency, \model{} exceeds RepoAgent (2.9) by 1.7 and RPG-Encoder (3.9) by 0.7, since \model{} carries a node's relation to its parent and siblings through the Deps and Data fields at generation, whereas isolated generation cannot guarantee consistency. The same margins appear in the LLM-as-a-Judge's scoring.

\begin{figure}[t]
  \centering
  \includegraphics[width=\columnwidth]{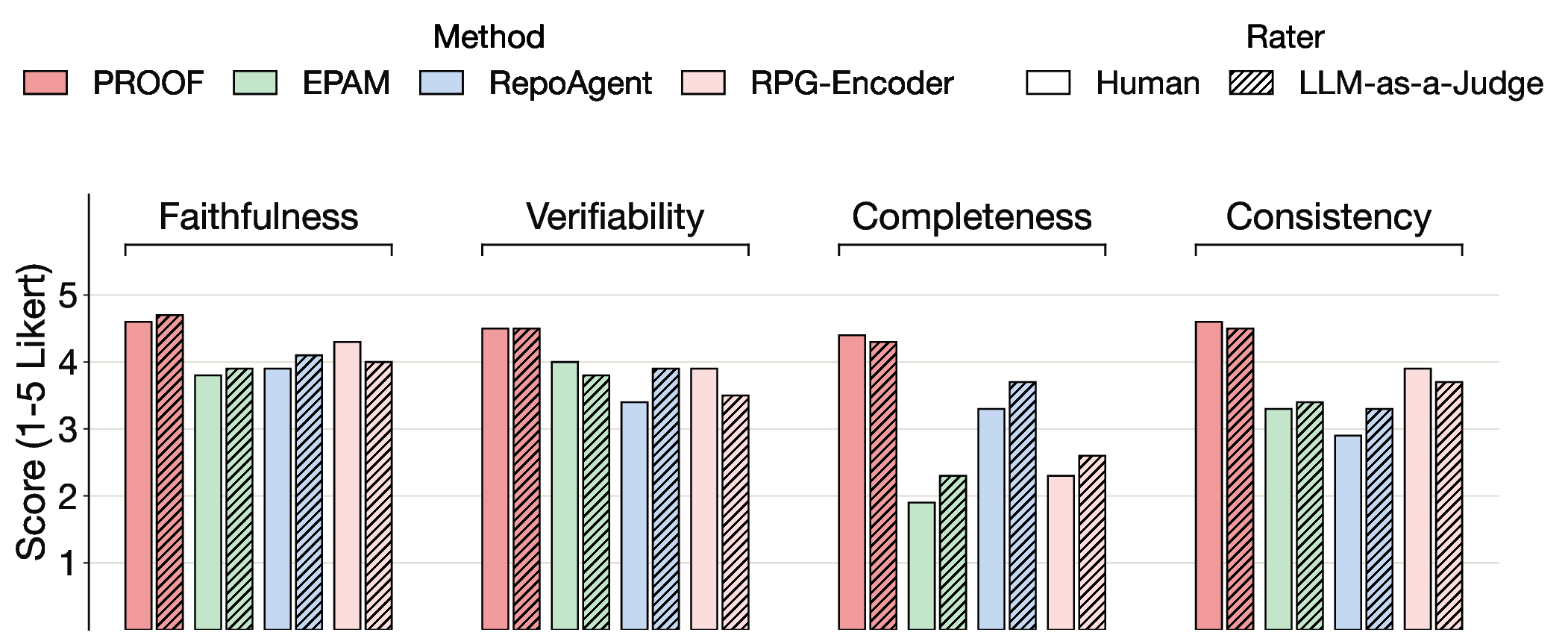}
  \caption{Spec Trustworthiness (Exp3): human (3-rater mean) and LLM-as-a-Judge (k=3 mean) scores across four content-quality dimensions.}
  \label{fig:spec-trustworthiness}
\end{figure}

\section{Conclusion}

Software now grows faster than engineers can maintain it at the code level, and applying large language models to repository-scale code is limited by context windows and complex dependency topologies, producing hallucinations and inconsistencies. We introduce a verified intermediate representation, the Spec, which lets developers locate, review, and drive code evolution through three stages: generation, verification, and update. Experiments show the Spec is accurate in structural topology, behavioral equivalence, and node-level content: its skeleton mirrors the codebase's organization, reconstructed code preserves original behavior, and node descriptions are validated as reliable by both human and model judges. The drift calibration mechanism ($\Delta_i$) further offers a way to detect and correct semantic drift during sustained evolution.

\clearpage
\section*{Limitations}

The applicability of \model{} depends on existing engineering infrastructure and underlying model capabilities. Verification relies on existing test suites to ensure objective behavioral equivalence. For codebases lacking sufficient test coverage, our method needs to incorporate complementary techniques like LLM-based test generation to complete the verification loop. Furthermore, \model{}'s practical effectiveness varies with the underlying LLM's semantic understanding of the target language. For low-resource languages or unseen private domain-specific languages (DSLs), our approach requires utilizing retrieval-augmented generation (RAG) or model fine-tuning to enhance the LLM's syntax comprehension. Finally, our current evaluation focuses on resolving static call graphs within individual repositories. Extending this framework to distributed architectures, such as systems relying on RPCs or message queues rather than static graphs, remains one of several potential directions for future research.

\bibliography{custom}

\appendix
\section{Spec Example}
\label{sec:appendix-spec-example}

Section~\ref{sec:hierarchy-design} defines the Object closure $v_{obj} = \langle \text{ClassDesc}, \text{DataState}, \text{Invariants}, M_{obj} \rangle$ and the Function closure $v_{func} = \langle \text{Desc}, \text{I/O}, \text{Meta}, \text{Deps}, \text{Data}, \text{Error} \rangle$. We illustrate both using three nodes actually generated by \model{} for Flask: the \texttt{Flask} class as an Object closure, its method \texttt{get\_send\_file\_max\_age} ($v_m \in M_{obj}$), and \texttt{\_prepare\_send\_file\_kwargs}, a module-level function declared outside any object whose Data field reads \texttt{ctx.app.get\_send\_file\_max\_age}, thereby connecting the two nodes across the object boundary within the same call graph.

\begin{figure*}[t]
  \centering
  \includegraphics[width=\textwidth]{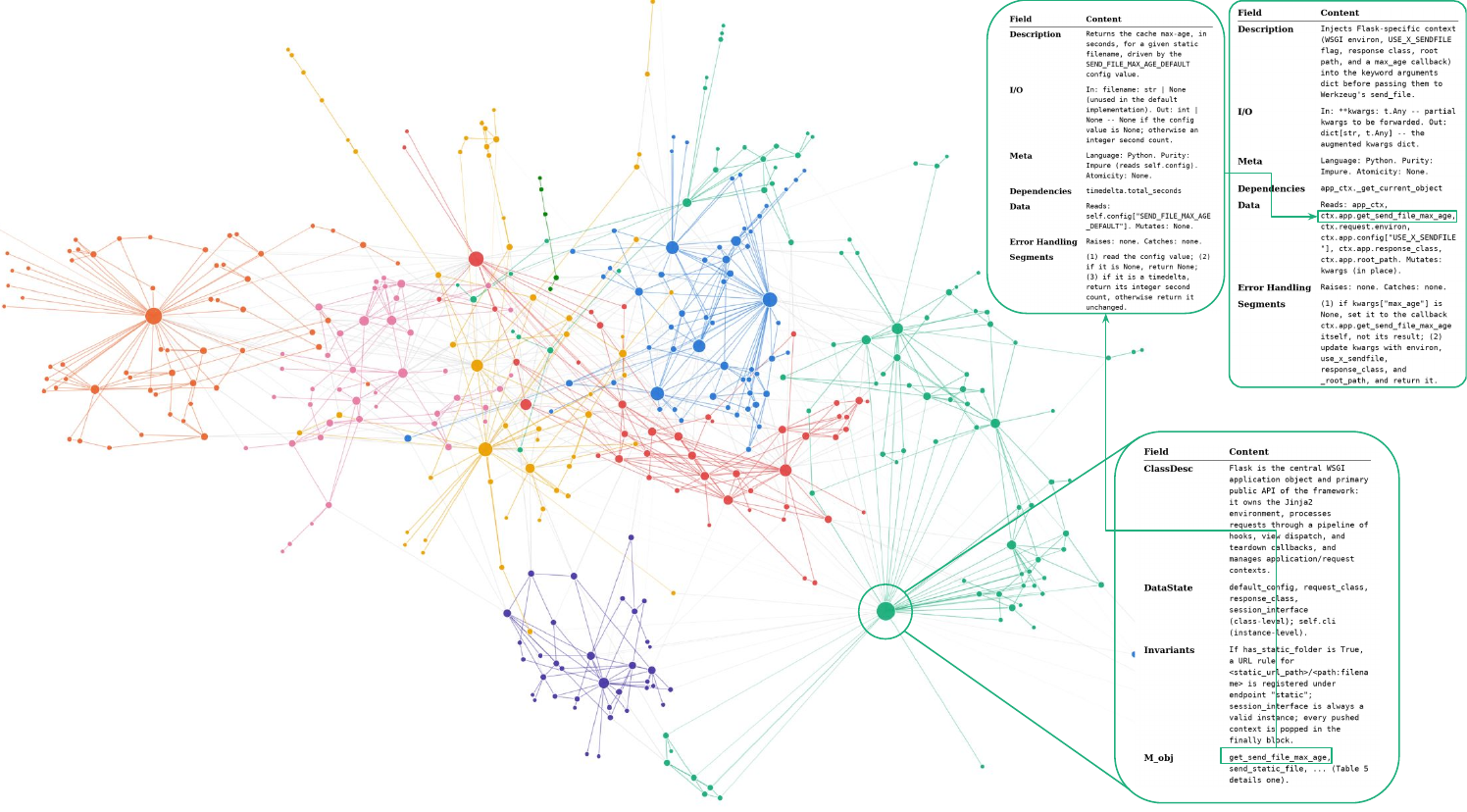}
  \caption{The \texttt{Flask} object located within the full call graph of the Flask repository (436 nodes, 1{,}016 edges), with its Object closure, the \texttt{get\_send\_file\_max\_age} method closure, and the \texttt{\_prepare\_send\_file\_kwargs} function closure that depends on it shown alongside the node they were generated from.}
  \label{fig:appendix-callgraph}
\end{figure*}

\begin{figure*}[t]
  \centering
  \includegraphics[width=\textwidth]{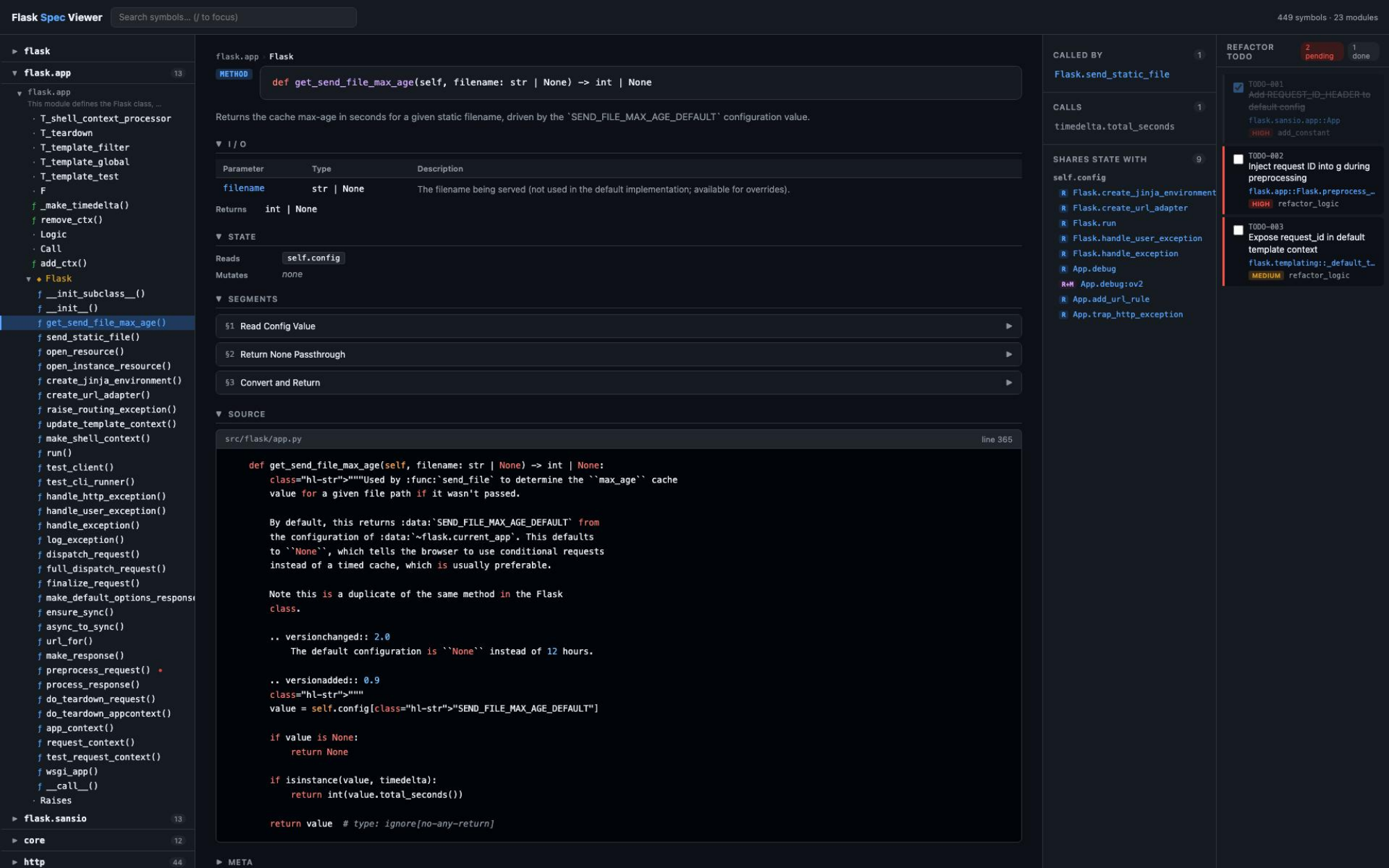}
  \caption{The Spec Viewer's interface for the same \texttt{get\_send\_file\_max\_age} node, laid out in three panels. \textbf{Left}: the codebase's call graph unfolded into a hierarchical, collapsible tree, mirroring the Spec's Project/Module/Object/Function levels rather than the raw file layout. \textbf{Center}: the selected node's Spec content (I/O, State, Segments) rendered alongside the source it maps to, so editing the Spec and watching the corresponding code update happen together in real time. \textbf{Right}: the Request panel, where an open maintenance request is decomposed into a checklist of concrete edits, each targeting one function; clicking an item jumps to that node for editing, and it is checked off once resolved. The request is complete once every listed function has been fixed, at which point the change is ready for testing.}
  \label{fig:appendix-viewer}
\end{figure*}

Figure~\ref{fig:appendix-callgraph} situates this triple within the full call graph of the Flask repository: the \texttt{Flask} object is one node among hundreds, and its closure content, along with that of its method and the standalone function that depends on it, is exactly what a developer or an LLM reads instead of the underlying source when working through \model{}. Figure~\ref{fig:appendix-viewer} shows the same \texttt{get\_send\_file\_max\_age} node in the interactive Spec Viewer, which renders every Function/Object closure as a navigable page and is laid out in three panels. The left panel exposes the codebase's full symbol hierarchy as a collapsible tree, folded module by module rather than by raw file path, mirroring the Spec's own Project/Module/Object/Function levels (Section~\ref{sec:hierarchy-design}). The center panel renders the selected node's closure (I/O, State, Segments) alongside the source it was generated from; a developer driving the edit through any LLM coding console (e.g., Claude Code, Antigravity, Codex) can update the Spec text and watch the corresponding code update beside it in real time, verifying the two stay consistent without leaving the page. The right panel resolves that node's Called-By/Calls/Shares-State-With edges, which correspond to the dependency edges shown in Figure~\ref{fig:appendix-callgraph}, and lists the maintenance request currently open against it as a checklist of per-node TODO items.

We illustrate this checklist with the concrete request shown in the figure: adding a per-request unique identifier for distributed tracing. The request decomposes into three node-level edits, each recording both a Spec-level change and its corresponding source change: (1) \texttt{App}'s \texttt{default\_config} gains a \texttt{REQUEST\_ID\_HEADER} constant defaulting to \texttt{None}, so the feature is opt-in; (2) \texttt{Flask.preprocess\_request} reads that header from the incoming request and stores it as \texttt{g.request\_id}, marking the first point in the request lifecycle where both \texttt{g} and \texttt{request} are available; (3) \texttt{\_default\_template\_ctx\_processor} exposes \texttt{g.request\_id} to templates under the key \texttt{request\_id}, following the pattern already used for \texttt{g}, \texttt{request}, and \texttt{config}. The first two edits are marked \texttt{high} priority, since the header name must exist in config before it can be read, and must be read before it can be exposed; the third, marked \texttt{medium}, is a usability addition layered on top once the first two hold. As shown in the figure's checklist, the developer checks off each edit on completion, giving a running view of how much of the request remains.

\section{Ablation Study}
\label{sec:appendix-ablation-tokens}

To examine the efficiency and performance of our method, we measure token consumption on Seaborn. Seaborn is the middle-scale repository among the three benchmark repositories used throughout the paper, covering the four systems compared in the main experiments: \model{}, RepoAgent, RPG-Encoder, and EPAM. We report consumption separately for the two stages of the pipeline: (1) \textbf{IR construction} (Code $\to$ intermediate representation) and (2) \textbf{reconstruction} (IR $\to$ Code), the latter broken down across three rounds ($k=3$) under the same repair protocol as Experiment 1 (Section~\ref{sec:exp-roundtrip}): Round 1 is the zero-repair, full-codebase generation, and Rounds 2--3 are test-failure-driven repair rounds.

Table~\ref{tab:ablation-ir} reports the size of each system's generated intermediate representation on Seaborn, along with the total token consumption for constructing this representation across the whole repository.

\begin{table}[h]
\centering
\begin{minipage}[b]{0.46\textwidth}
\centering
\small
\begin{tabular}{lrr}
\toprule
System & IR words & Tokens \\
\midrule
RPG-Encoder & 18,206 & 3.32M \\
\model{} & 126,743 & 4.32M \\
RepoAgent & 549,948 & 6.03M \\
EPAM & 237,889 & 9.61M \\
\bottomrule
\end{tabular}
\captionof{table}{Size of each system's generated IR on Seaborn and IR construction token consumption.}
\label{tab:ablation-ir}
\end{minipage}
\hfill
\begin{minipage}[b]{0.5\textwidth}
\centering
\includegraphics[width=\linewidth]{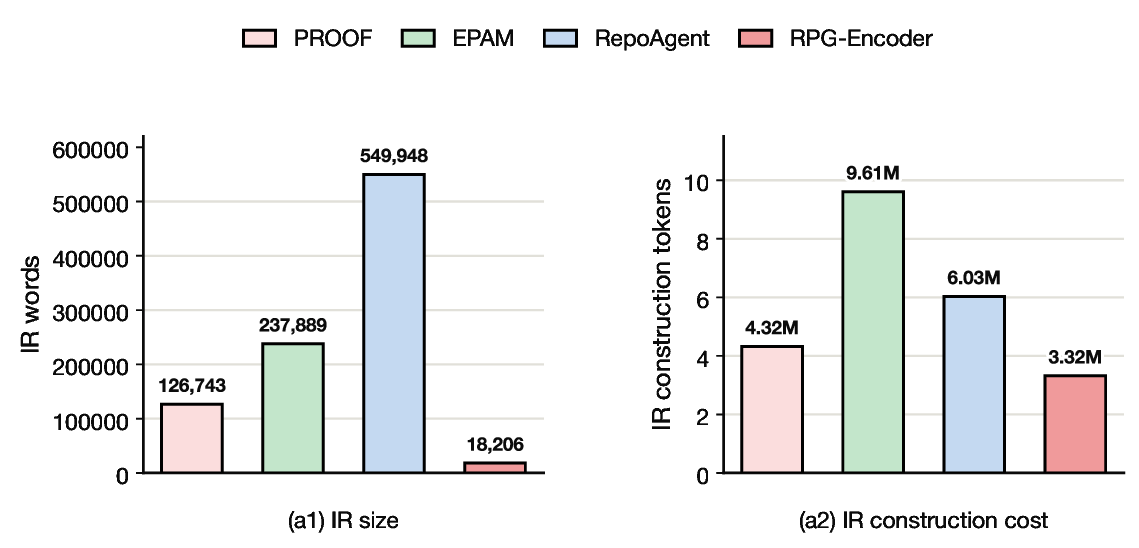}
\captionof{figure}{IR words (a1) and IR construction tokens (a2) per system on Seaborn, drawn as two small-multiple panels rather than one dual-axis chart since the two measures are different units at very different magnitudes.}
\label{fig:ablation-ir}
\end{minipage}
\end{table}

\model{}'s IR construction consumes 4.32M tokens on Seaborn, 1.30$\times$ RPG-Encoder's 3.32M. RepoAgent, which generates free-form per-entity documentation without batching, consumes 6.03M tokens. EPAM, lacking a tightly-constrained topological structure, likely consumes more tokens navigating complex dependency relations.

Building on this, we further measure consumption at the reconstruction stage. Table~\ref{tab:ablation-recon-tokens} reports the total token consumption for reconstructing Seaborn's codebase from each system's representation, across all three rounds (Round 1 is the zero-repair, full-codebase generation; Rounds 2--3 are test-failure-driven repair rounds).

\begin{table}[h]
\centering
\begin{minipage}[b]{0.46\textwidth}
\centering
\small
\begin{tabular}{lrr}
\toprule
System & Round 1 & Total \\
\midrule
RPG-Encoder & 3.65M & 6.12M \\
\model{} & 5.29M & 5.49M \\
RepoAgent & 4.75M & 6.59M \\
EPAM & 12.01M & 14.20M \\
\bottomrule
\end{tabular}
\captionof{table}{Reconstruction-stage token consumption on Seaborn, $k=3$.}
\label{tab:ablation-recon-tokens}
\end{minipage}
\hfill
\begin{minipage}[b]{0.5\textwidth}
\centering
\includegraphics[width=\linewidth]{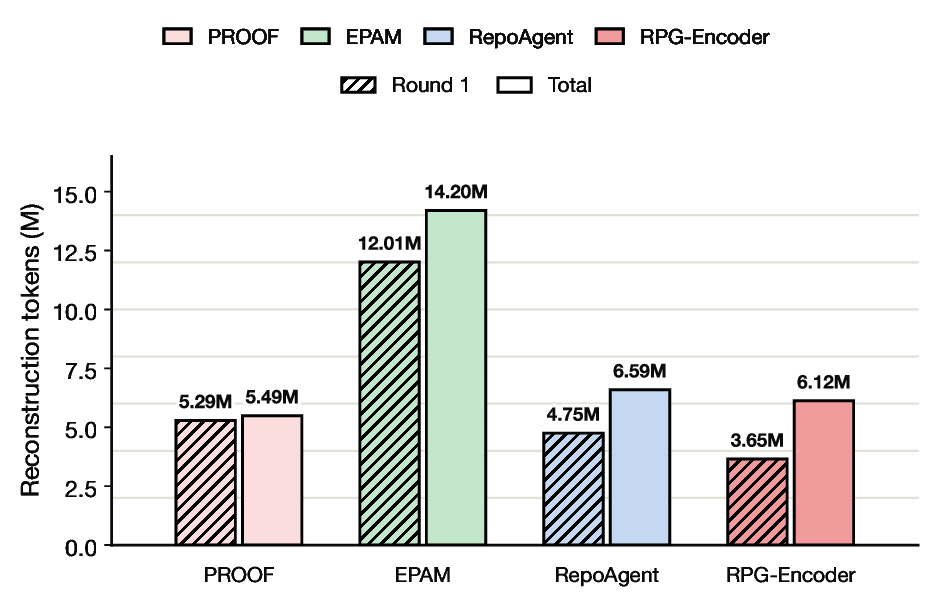}
\captionof{figure}{Round 1 (hatched) versus total (solid) reconstruction tokens per system on Seaborn, $k=3$.}
\label{fig:ablation-recon-tokens}
\end{minipage}
\end{table}

\model{} requires a substantial initial investment in Round 1 (5.29M tokens), exceeding both RPG-Encoder (3.65M) and RepoAgent (4.75M), though remaining more efficient than EPAM. We attribute this to the structured six-field contract, which produces the most information-dense representation per node but also carries the highest per-token cost when consumed as generation input. This upfront cost, however, is fully offset during the repair phase. Because \model{} converges rapidly, its subsequent repair rounds require minimal additional tokens, bringing its total consumption (5.49M) below RPG-Encoder's (6.12M). In contrast, RepoAgent (6.59M) and EPAM (14.20M) both struggle to achieve full convergence, continuing to incur substantial repair costs in subsequent rounds without fully resolving. \model{}'s advantage in total reconstruction cost is therefore conditional: because it converges faster, spending more tokens in the first round pays off.

Beyond evaluating the token efficiency of IR construction and reconstruction, it is crucial to determine whether \model{}'s performance is overly reliant on the reasoning capabilities of a specific backbone model. Large language models exhibit varying degrees of proficiency in instruction following, context management, and zero-shot code synthesis. To ensure our framework's robustness and demonstrate that the observed advantage is not merely an artifact of a single powerful LLM, we re-run the Generation and Verification pipeline on the Seaborn repository using three distinct backbone models and report the resulting Pass@1 metrics. For this ablation study, we interface with each model through its native environment to maximize expected performance: Claude Console for Claude 5 Sonnet, Codex for GPT-5.5 Medium, and Antigravity for Gemini 3.1 Pro.

\begin{table}[t]
\centering
\small
\begin{tabular}{lr}
\toprule
Backbone & Pass@1 (Seaborn) \\
\midrule
Claude 5 Sonnet & 83.5\% \\
Gemini 3.1 Pro & 72.1\% \\
GPT-5.5 Medium & 55.7\% \\
\bottomrule
\end{tabular}
\caption{\model{}'s Seaborn Pass@1 under different backbone LLMs.}
\label{tab:ablation-backbone}
\end{table}

As detailed in Table~\ref{tab:ablation-backbone}, the choice of backbone LLM undeniably influences the final reconstruction quality, with Pass@1 scores scaling predictably with the underlying model's general capabilities (dropping from 83.5\% to 72.1\% and finally to 55.7\%). This variance confirms that the framework still benefits significantly from advanced base models, as reconstruction is not entirely backbone-independent. However, the most critical finding emerges when comparing across methods rather than across models: even when powered by the weakest backbone in our test suite (GPT-5.5 Medium at 55.7\%), \model{} still substantially exceeds the highest-performing baseline powered by the most capable model (RepoAgent using Claude 5 Sonnet, which achieved only 29.95\% as shown in Table~\ref{tab:roundtrip-results}). This absolute margin of over 25 percentage points isolates the impact of the framework's core design. It provides strong empirical evidence that \model{}'s advantage over existing baselines is fundamentally attributable to the method itself, specifically the hierarchical, dependency-aware Spec generation process and the rigorous constraints of the structured six-field contract, rather than merely piggybacking on the raw computational power or context window of any single backbone model.

\section{Human Evaluation Instructions}
\label{sec:appendix-human-eval}

This is the rating form used by human raters for the Interpretability experiment (Section~\ref{sec:exp-trust}): a 1--5 Likert scale with anchors, scored per node description.

\paragraph{Rater instructions.}
\begin{itemize}[leftmargin=0.2in, itemsep=0pt, parsep=0pt, topsep=3pt]
  \item Each rating covers one node's description. Raters do not know which system produced it (\model{} / RepoAgent / RPG-Encoder / EPAM, order shuffled and labels hidden, randomly numbered as Description A/B/C/D).
  \item Each node is rated alongside its source code (for Faithfulness / Verifiability / Completeness) and its parent/sibling node descriptions (for Consistency).
  \item The four dimensions are scored independently, 1--5, half-points allowed (e.g., 3.5).
\end{itemize}

\begin{table*}[t]
\centering
\footnotesize
\begin{adjustbox}{max width=\textwidth}
\begin{tabular}{p{2.1cm}p{3.1cm}p{3.1cm}p{3.1cm}p{3.1cm}p{3.1cm}}
\toprule
Dimension & 1 & 2 & 3 & 4 & 5 \\
\midrule
\textbf{Faithfulness} & Description seriously diverges from or fabricates the code's behavior & Broadly correct, but multiple inaccurate or misleading statements & Mostly accurate, with one or two detail-level deviations & Accurate, with only minor flaws that do not affect understanding & Fully accurate, matches the code's actual behavior exactly, no fabrication \\
\textbf{Verifiability} & Entirely generic phrasing (e.g., ``processes data'', ``executes logic''), no claims checkable against the code & Mostly generic, only a few specific, checkable statements & About half the content is specific and checkable, still many vague statements & Most statements are specific and clearly pointed, individually verifiable & Every statement is specific and checkable, traceable line-by-line to the code \\
\textbf{Completeness} & Covers almost none of I/O, side effects, error handling, dependencies & Covers only a small fraction of key information, omitting content critical to understanding/modifying the function & Covers most key information, but one or two items are clearly missing & Key information is mostly complete, only minor (atypical edge-case) information missing & I/O, side effects, error conditions, dependencies, and all key information fully covered \\
\textbf{Consistency} & Clearly contradicts the parent or sibling node (e.g., inconsistent interface/data-flow description) & Multiple noticeable inconsistencies or awkward transitions & Broadly self-consistent, with one or two minor wording/granularity mismatches & Basically consistent with context, only very minor phrasing differences & Fully self-consistent with parent/sibling nodes, uniform terminology, granularity, and boundary description \\
\bottomrule
\end{tabular}
\end{adjustbox}
\caption{Human evaluation rubric: 1--5 Likert scale with anchors, one form filled per node per anonymized description.}
\label{tab:appendix-human-eval-rubric}
\end{table*}

\end{document}